\documentstyle[fleqn,epsf,rotate]{elsart}

\begin{document}
\begin{frontmatter}
\title{Crossover scaling functions and an \\
extended minimal subtraction scheme\thanksref{talk}}
\thanks[talk]{Expanded version of an invited talk presented at the 
StatPhys-Taipei-1995 (Taipei, July 1995).}
\author{Erwin Frey\thanksref{DFG}}
\address{Institut f\"ur Theoretische Physik,
Physik-Department der Technischen Universit\"at M\"unchen,
James-Franck-Stra\ss e, D-85747 Garching, Germany}
\thanks[DFG]{Supported by a Habilitation grant from the Deutsche 
Forschungsgemeinschaft.}
\begin{abstract}
A field theoretic renormalization group method is presented which is capable of dealing
with crossover problems associated with a change in the upper critical dimension. The
method leads to flow functions for the parameters and coupling constants of the model
which depend on the set of parameters which characterize the fixed point landscape of
the underlying problem. Similar to Nelson's trajectory integral method any vertex
function can be expressed as a line integral along a renormalization group trajectory,
which in the field theoretic formulation are given by the characteristics of the
corresponding Callan-Symanzik equation. The field theoretic renormalization
automatically leads to a separation of the regular and singular parts of all crossover
scaling functions.

The method is exemplified for the crossover problem in magnetic phase transitions,
percolation problems and quantum phase transitions. The broad applicability of the
method is emphasized.
\end{abstract}
\end{frontmatter}

\section{Introduction}

Systems in the vicinity of a critical point show scale invariance~\cite{ST95}. Quantities like the
susceptibility obey a homogeneity property, which is characterized by critical exponents
and scaling functions. Within the framewrk of renormalization group theory it can be shown
that this scaling property is closely related to the existence of a fixed point of the
renormalization group transformation. In the simplest case the critical behavior of a
particular problem is dominated by just one fixed point. Quite frequently, however, one
encounters physical systems which can exhibit different scaling behavior in different
asymptotic regimes. In those cases it becomes important to study  the crossover from one 
type of critical behavior to another. Examples for such crossover problems are 
bicritical and tricritical points (in general multicritical points), magnetic systems with
anisotropic interactions such as anisotropies in the exchange interaction as well as
dipolar interaction, finite size problems, and in general any problem which is
characterized by some length scales in addition to the correlation length $\xi$.

The theoretical investigations of those crossover phenomena started with a scaling theory
by Riedel and Wegner~\cite{RW69}, and the extended scaling hypothesis of Pfeuty, Fisher 
and Jasnow~\cite{PFJ74}.
Nelson has presented the first renormalization-group formalism for calculating the 
crossover scaling functions associated with such situations, the so called trajectory
integral method~\cite{Nel75}. In this paper we will present a field theoretic version of
Nelson's trajectory integral method and exemplify the method by several problems from
various fields of critical phenomena. The method is based on the work by Amit and Goldschmidt
~\cite{AG78} who proposed a ``generalized minimal subtraction method'', which has subsequently been
applied to various crossover problems~\cite{Law81,Fre88,TS92}. Recently the method has been
extended to deal with situations, where the crossover is accompanied by a change in the upper
critical dimension~\cite{Fre90,FST94}.

In many crossover problems one encounters a situation where besides the correlation
length $\xi$ there is a second length scale, characterized by the anisotropy scale $g$.
Out of many examples we would like to discuss three quite characteristic cases.

(1) The critical behavior of a {\it uniaxial dipolar ferromagnet} is described by the
effective Ginzburg-Landau-Wilson (GLW) free energy functional~\cite{Fre90}
\begin{equation}
  {\cal H} = \half \int_{\bf k} 
  \left[ r+k^2+g^2{q^2 \over p^2} \right] \phi^\alpha ({\bf k}) \phi^\alpha ({-\bf k})
   + {u \over 4!} \int_{\bf x} [\phi^\alpha ({\bf x}) \phi^\alpha ({\bf x}) ]^2 \, ,
\label{1}
\end{equation}
where $g$ measures the strength of the dipolar interaction relative to the exchange
interaction, and the wave vector ${\bf k}$ has been decomposed into a component $q$ along the
uniaxial direction and the remaining components ${\bf p}$.

(2) The crossover from isotropic to directed percolation can be described by a type of 
{\it Reggeon field theory} with the action~\cite{Car80,Ben84}
\begin{eqnarray}
  {\cal S} = &&\int_{\bf x} \int_t 
  {\tilde \phi} ({\bf x},t )
  \left[ r -\nabla^2 - {1 \over c^2} {\partial^2 \over \partial t^2} + {2 g \over c}
         {\partial \over \partial t} 
  \right]
  \phi ({\bf x},t ) \nonumber \\
   &&+ {u \over 2}   \int_{\bf x} \int_t 
  \left[  {\tilde \phi}^2 ({\bf x},t )  \phi ({\bf x},t ) - 
          {\tilde \phi}   ({\bf x},t )  \phi^2 ({\bf x},t )
  \right] \, .
\label{2}
\end{eqnarray}

(3) Multicritical behavior near the superfluid to Mott insulator quantum phase 
transition in Bose lattice models are described by the effective action|cite{FGW89}
\begin{equation}
  {\cal S} = \half \int_{{\bf k}, \omega} 
  \left( r + k^2 + 2 g i \omega + \omega ^2 \right) \mid \phi ({\bf k}, \omega) \mid^2
  + {u \over 4} \int_{{\bf x}, t}  \mid \phi ({\bf x}, t) \mid^4 \, , 
\label{3}
\end{equation}
where the field $\phi$ serves as an order parameter for superfluidity and the parameter 
$g$ (which vanishes at the multicritical point) is a measure of the absence of 
particle-hole symmetry.

All these phenomena have in common the characteristic 
that the anisotropy scale appears in the harmonic
part of the effective functional of the model, and that asymptotically there is a change
in the upper critical dimension (usually a reduction). Hence there are not only
different critical points complicating the renormalization group analysis, but in
addition those fixed points are characterized by different upper critical dimensions.
The latter invalidates conventional methods such as $\varepsilon$-expansion since there is
no fixed upper critical dimension to expand around. On the contrary the crossover may be
even viewed as a continuous variation of the upper critical dimension.

The paper is organized as follows. In the following section we introduce an extended
minimal subtraction procedure which is capable of describing crossover phenomena from one
fixed point to another accompanied by a change in the upper critical dimension. In section
3 we exemplify the method for the three models mentioned above. Finally, we summarize our
results in section 4 and give an outlook concerning the broad applicability of the method.

\section{The method of generalized minimal subtraction}

Crossover phenomena are characterized by the presence of at least one additional 
relevant length scale $g$ besides the correlation length $\xi \propto r^{-\nu}$. 
Frequently it is given by an anisotropy or ``mass'' parameter describing the variation 
from one scaling region to another, or in renormalization group language, from a primary 
fixed point to a secondary fixed point\footnote{In general the scenario may be even more 
complicated. Depending on the number of length scales present in the problem there may by 
more than two fixed points with a cascade of various crossover phenomena.}. We have 
listed some examples in the introduction. In order to describe this crossover from one 
critical behavior to another Riedel and Wegner~\cite{RW69} 
have formulated a scaling theory, and 
later Pfeuty, Fisher and Jasnow~\cite{PFJ74} proposed an extended scaling hypothesis for 
crossover 
systems. Let $r$ be the reduced temperature and $g$ be the scaling field characterizing 
the additional length scale in the problem, then the crossover scaling Ansatz, e.g. for the 
susceptibility, reads
\begin{equation}
  \chi(r,g) = r^{-\gamma} f (g/r^\phi)
\label{4}
\end{equation}
near the multicritical point $(r,g) = (0,0)$. Here $\gamma$ is the susceptibility exponent
{\it at} the primary fixed point, while $\phi$ is the crossover exponent. In order to describe
the change of the critical indices induced by the crossover from the primary to the
secondary fixed point, one assumes that the crossover scaling function $f(x)$ becomes
singular at $x_c$
\begin{equation}
  f(x) \propto (x_c - x)^{-{\dot \gamma}}
\label{5}
\end{equation}
as $r \rightarrow r_c (g)$ approaches criticality for fixed $g$. As is obvious from the
above description of the phenomenological crossover scaling theory, it has the 
disadvantage that the crossover in the critical indices is incorporated in the crossover 
scaling function in a rather singular way. It would be much more convenient to have a 
formulation of the crossover problem, where the crossover of the exponents is 
incorporated in the power law prefactor $r^{-\gamma}$ by $\gamma$ becoming an effective 
exponent depending on the crossover scaling variable $ x = g/r^\phi$. This would leave us 
with a regular scaling function $f^\prime (x)$. As will become clear later the
field theoretic method proposed in this section, will satisfy the latter requirement. 

The first quantitative theory for crossover scaling functions based on the
renormalization group theory was given by Nelson~\cite{Nel75}. 
He introduced a formalism which
expresses the free energy as a trajectory integral along the renormalization group flow
lines. The method proposed below will be similar in spirit to this trajectory integral
method, but will have considerable technical advantages due to the explicit separation of
singular and regular terms in the crossover scaling function.

Now we describe a field theoretic method for the analysis of crossover phenomena. The 
technical difficulty one encounters here is that both the ultraviolet (UV) and infrared
(IR) singularities will 
differ in the two distinct scaling regimes. Early work on
formulating crossover phenomena in terms of field theoretic methods, tried to parallel
the formulation using phenomenological scaling theories. In this 
``traditional'' approach to crossover problems, one would compute the critical 
exponents at one of the stable fixed points; all the crossover
features would then be contained in the accompanying scaling function as 
corrections to this scaling behavior. However, in general a calculation to high
order in the perturbation expansion would be required in order to achieve a 
satisfactory description of the entire crossover region. Of course, using an 
($\varepsilon = d_c - d$) expansion with respect to either of the fixed points
renders the other one completely inaccessible, if their upper critical
dimensions do not coincide. Therefore Amit and Goldschmidt's idea~\cite{AG78,Ami84} to
incorporate the crossover features already in the exponent functions has proven
much more successful than treating the problem on the basis of scaling
functions. The essential prescription one has to bear in mind is that the
renormalization constants are not solely functions of the anharmonic coupling,
but necessarily also of the additional ``mass'' or anisotropy parameter $g_0$
describing the interplay between the two different scaling regimes~\cite{AG78}. 
For a consistent treatment of the entire crossover region, 
one thus has to assure that the UV singularities are absorbed into the Z factors for any
arbitrary value of $g_0$, including $g_0 \rightarrow \infty$. This is not a
trivial prescription, as usual the $1 / (d_c-d)$ poles will be altered in the
different scaling regimes. For the situation that we have in mind, even the
value of the upper critical dimension is bound to change as the crossover takes
place, in contrast to previously studied cases \cite{Law81,Fre88,TS92}.
However, we shall demonstrate that with the above stated so-called
``generalized subtraction scheme'' this change in the upper critical dimension
may be incorporated into the usual formalism without any drastic changes, provided
one refrains from any $\varepsilon$ expansion about $d_c$. The 
perturbation series is then an expansion with respect to the effective coupling
$v$ to be introduced later, which is not an a-priori small parameter, and the
perturbation expansion is uncontrolled in this sense. If higher
orders of the perturbation expansion were known, one could substantially 
refine the theory by a Borel resummation procedure. For a more detailed
discussion of the question in which cases one may dispense with a $(d_c-d)$
expansion, we refer to work of Schloms and Dohm \cite{Doh89}.

We remark that the somewhat misleading term
``generalized minimal subtraction scheme'' stems from the fact that in the framework of
an $\varepsilon$ expansion this corresponds to adding logarithms of $g_0$, which
are finite in the limit $\varepsilon \rightarrow 0$, to the $Z$ factors. This is the
procedure proposed in the original work by Amit and Goldschmidt~\cite{AG78}. But it
works only if the upper critical dimension of the primary and secondary fixed point
are equal. Otherwise the method by  Amit and Goldschmidt has to be extended. 
There are several ways to implement the requirement that the vertex functions become
finite in both limits $g_0 \rightarrow 0$ and $g_0 \rightarrow \infty$ after 
reparameterizing the theory using renormalization factors. We will explicitly use two of
them. Firstly, for the uniaxial dipolar ferromagnet, discussed in section 3.1,
we will use the prescription of Amit and Goldschmidt, extended in the following
way. Upon performing an $\varepsilon$-expansion around the upper critical dimension
of the primary fixed point, one gets to one loop order 
renormalization factors of the structure
$Z = 1 + {a \over \varepsilon} + b \ln g_0$. We reexponentiate those logarithms of $g$
such as to map the result one would obtain by doing an expansion of the model
for $g_0 \rightarrow \infty$ around the upper critical dimension of the secondary
fixed point. This prescription allows for an analytic treatment of the flow equations.
Secondly, we will determine the renormalization factors by requiring that
it contains all the singularities for both limits $g_0=0$ and $g_0=\infty$, but without
performing any expansion around any of the critical dimensions. A (minor) price to be 
paid is that for the flow equations and related quantities only numerical solutions are 
accessible, and merely the limiting cases of $g_0 = 0$ and $g_0 \rightarrow \infty$, 
respectively, allow for an analytical investigation. The latter method is quite close
to the method of normalization conditions~\cite{Ami84}, 
but is minimal in that it contains just the
poles in ${1 \over \varepsilon}$ for both limits $g_0 = 0$ and $g_0 \rightarrow \infty$.
Here $\varepsilon$ refers to the distance to the upper critical dimension of the primary and
secondary fixed point, respectively.

\subsection{Flow equations}

Although the following discussion of the flow equations can be presented more generally,
it is convenient to have the specific set of effective free energy functionals and
actions in mind, which we have presented in the introduction. Therefore, we assume that
all of Wilson's flow functions characterizing the crossover in the fixed points and the
corresponding critical exponents depend only on two parameters, 
$\zeta = \zeta (u, g)$.

The renormalization group equations relate the values of an arbitrary vertex function
at one point in parameter space to a transformed point. Conceptually the renormalization
group equations are obtained by observing that all bare vertex functions $\Gamma^0$ are
independent of the arbitrary renormalization scale $\mu$
\begin{equation}
  \mu {\partial \over \partial \mu} \big\vert_0 \Gamma^0 ({\bf a}, g, u; {\bf q})
  = 0 \, ,
\label{6}
\end{equation} 
where ${\bf a}$ is a set of parameters characterizing the effective free energy
functional under consideration, and $\big\vert_0$ indicates that all derivatives have
to be taken at fixed values of the bare parameters. The quantity $g$ is the second 
length scale of the problem besides the correlation length $\xi$, and $u$ is the coupling
constant of the nonlinearity. Wilson's flow functions are defined by
\begin{equation}
  \zeta_x = \mu {\partial \over \partial \mu} \big\vert_0 \ln (x/x_0) \, ,
\label{7}
\end{equation}
where $x$ stands for the parameters ${\bf a}$, the anisotropy scale $g$ and the coupling
constant $u$. Furthermore we introduce the $\zeta$-function for the renormalization of
the vertex function $\Gamma$
\begin{equation}
  \zeta_\Gamma = \mu {\partial \over \partial \mu} \big\vert_0 \ln Z_\Gamma \, ,
\label{8}
\end{equation}
where $\Gamma^0 = Z_\Gamma \Gamma$. Then one derives the following partial differential
equation
\begin{equation}
     \left[     \mu {\partial \over \partial \mu} + 
             \sum_i \zeta_{a_i} a_i {\partial \over \partial a_i} + 
             \zeta_g g {\partial \over \partial g} +
             \zeta_u u {\partial \over \partial u} + \zeta_\Gamma \right] 
                              \Gamma (\mu,{\bf a},g,u;{\bf q}) = 0 
\label{9}
\end{equation}
for the renormalized vertex function.
As a consequence of the generalized renormalization scheme, all the flow functions are
functions of the coupling constant $u$ as well as the anisotropy scale $g$. This leads
to renormalization group trajectories in a two-dimensional parameter space, the 
$(g,u)$-plane. Within this plane there are usually several fixed points, where two of
them are of particular importance. These are the primary fixed point $P$ located at $g=0$
and the secondary fixed point $S$ located at $g=\infty$. The primary fixed point turns
out to be infrared stable for $g=0$ only, whereas all the renormalization group flow 
tends to the secondary fixed point $S$ if the anisotropy scale $g$ is nonzero.
The renormalization group flow within Wilson's elimination procedure is generated by the
introduction of a spatial rescaling factor $b$. In the present field theoretic
framework, a renormalization group trajectory is given in terms  of the characteristics
of the Callan-Symanzik equation, Eq.(\ref{9}). The characteristics are defined as the
solution of the following first order differential equations
\begin{equation}
     \ell {d x(\ell) \over d \ell} = \zeta_x(\ell) x(\ell) \, ,
\label{10}
\end{equation}
with the initial conditions $x(1) = x$, namely
$x(\ell) = x \exp {\int_1^\ell \zeta_x(\ell^\prime) d \ell'/ \ell'}$. 
If we denote the engineering dimension of the vertex function $\Gamma$ by $d_\Gamma$ and
define the dimensionless (renormalized) vertex function ${\hat \Gamma}$ by
\begin{equation}
   \Gamma (\mu, \{ x \}; {\bf q}, \omega) 
   = \mu^{d_\Gamma} {\hat \Gamma} (\{ x \}; {\bf q}/\mu) \, ,
\label{12}
\end{equation}
the solution of Eq.(\ref{9}) reads
\begin{equation}
     \Gamma (\mu,\{ x \};{\bf q}) =                 
      (\mu \ell)^{d_\Gamma}  
      \exp \left\{ \int_1^\ell \zeta_\Gamma (\ell') {d \ell' \over \ell'} \right\}
      {\hat \Gamma} \left( {\bf a} (\ell), g(\ell), v(\ell); {{\bf q} \over \mu \ell}
             \right) \, . 
\label{13}
\end{equation}
Here we have introduced an effective anharmonic coupling $v$, which is some combination
of the nonlinearity $u$ and the anisotropy scale $g$, where the detailed form
depends on the particular problem one is considering. The introduction of this effective
coupling constant is necessitated by the fact that one should work with coupling 
constants whose fixed point values are finite in both limits, 
$g\rightarrow 0$ and $g \rightarrow \infty$. The representation (\ref{13}) of the
vertex function in terms of the characteristics of the parameters and coupling constants
constitutes an explicit expression for the desired crossover scaling
function\footnote{Here we do not consider the case of composite-operator 
vertex functions, which have to be
renormalized additively. For an extension of the method to include these additively
renormalized quantities we refer the reader in particular to Ref.~\cite{Fre88}}.

Compared to the trajectory integral method of Nelson~\cite{Nel75} the above formalism has the
advantage that the singular and the regular part of the crossover scaling functions are
explicitly separated. This is simply a consequence of the field theoretic renormalization 
procedure.

\subsection{Flow diagram, fixed points and exponents}

In this subsection we give a qualitative discussion of the renormalization group flow in
the $({\bar g}, v)$-plane, where ${\bar g} = g/(1+g)$ and $v$ are appropriately defined
effective coupling constants, which are finite in the isotropic ($g=0$) and anisotropic 
($g=\infty$) limit. Usually the topology of the flow diagram is determined by the
presence of four fixed points, the primary and the secondary fixed points, and two
Gaussian fixed points with $v=0$ at $g=0$ and $g=\infty$. The primary fixed
point is infrared stable only if the anisotropy parameter equals zero. Otherwise the
only infrared stable fixed point is the secondary fixed point. All other fixed points
are unstable. The flow diagram is
separated into two regions by a separatrix, which is given by the trajectory originating 
from the primary fixed point and terminating in the secondary fixed point. This
renormalization group trajectory describes the universal crossover from primary to
secondary critical behavior. 

The flow diagram can be investigated quantitatively upon solving the 
flow equation of the running coupling $v(\ell)$ 
\begin{equation}
     \ell {d v(\ell) \over d \ell} = \beta_v(\ell) \, , 
\label{14}
\end{equation}
where the $\beta$ function is defined by
$\beta_v = \mu {\partial \over \partial \mu} \bigg \vert_0 v$.
In the flow equations above, the parameter $\ell$ may be considered as
describing the effect of a scaling transformation upon the system. Roughly speaking
$\ln \ell$ corresponds to the Wilson rescaling factor $b$. Obviously,
the theory becomes scale-invariant when a fixed-point $v^*$, to be obtained as
a zero of the $\beta$ function, $\beta_v(v^*) = 0$, is approached. The properties of 
the vertex function in the vicinity of the fixed
point will yield the correct asymptotic behavior, if the latter is
infrared-stable. Upon defining the fixed point values of the flow functions by
$\zeta_a^* = \zeta_a(v = v^*)$, also called the anomalous dimensions of the parameters 
$a$, we find 
\begin{equation}
     \Gamma (\mu, \{ x \};{\bf q}) = 
     \mu^{d_\Gamma} \ell^{d_\Gamma + \zeta_\Gamma^*}
      {\hat \Gamma} \left( \{ a_i \ell^{\zeta_{a_i}^*} \}, g \ell^{\zeta_g^*}, v^*;
      {{\bf q} \over \mu \ell} \right) \, .
\label{16}
\end{equation}
Using the matching condition $\ell = q / \mu$, one arrives at the following scaling form
\begin{equation}
     \Gamma (\mu, r, g, v^* ; {\bf q}) \propto 
      q^{2 - \eta}
      {\hat \Gamma}_{11} \left( {r \over (q / \mu)^{1 / \nu}}, g (q/\mu)^{\zeta_g^*}, 
      v^* ;  1 \right) \, ,
\label{17}
\end{equation}
where we have specialized to the simple case that the set of parameters $a_i$
reduces to just one parameter, namely the reduced temperature $r$.
We defined two independent critical exponents according to
$\eta = - \zeta_\Gamma^*$ and  $\nu  = - {1 \over \zeta_r^*}$.
The crossover exponent $\phi$ can be read off from the homogeneity relation for the vertex
function upon using the matching condition $r(\ell) \propto r \ell^{\zeta_r^*} = 1$
\begin{equation}
     \Gamma (\mu, r, g, v ; {\bf q}) \propto 
      r^\gamma
      {\hat \Gamma} \left( 1, g/r^\phi, v^*,;
      {q \over \mu} r^{- \nu} \right) \, ,
\label{19}
\end{equation}
where $\gamma = \nu (2 - \eta)$ and $\phi = \zeta_g^* / \zeta_r^*$.
Note that all of the above exponents correspond to the exponents at the secondary, and
asymptotically stable, fixed point. If one could neglect the variation of the scaling
functions on the flow of the parameters on the right hand side of Eq.(\ref{13}), 
the effective exponents of the problems would simply be given by equations like 
$\eta = - \zeta_\Gamma^*$, where 
the fixed point values of the Wilson flow functions are replaced by their values along the
renormalization group trajectory. Such an approximation is quite frequently referred to
as a {\it ``renormalized mean field'' approximation}.

\section{Model systems with dimensional crossover}

\subsection{Uniaxial dipolar ferromagnet}

The influence of the dipolar interactions on the critical behavior of magnetic systems
with isotropic and uniaxial exchange interaction is quite different, especially if one
takes into account the dipole-dipole interaction present in any real ferromagnetic
material. For isotropic ferromagnets the dipolar interaction leads only to a 
slight modification of the critical exponents. Uniaxial dipolar ferromagnets, however, 
show classical behavior with logarithmic corrections in three 
dimensions~\cite{LK69,Aha73,BZ76}, which in 
the language of renormalization group theory means that the upper critical dimension for 
the asymptotic critical behavior is shifted to $d_c=3$.

The existence of logarithmic corrections was verified experimentally for a number of
uniaxial ferromagnetic substances~\cite{AKG75,FKS82}. However, these experiments were 
performed in
regions of reduced temperature, where departures from the asymptotic behavior are 
expected and are  indeed observed. In particular, one finds a maximum in the effective 
exponent of the susceptibility~\cite{FKS82}. Here we study this crossover by a the
extended minimal subtraction described in the preceding section. This method
allows the calculation of the complete flow of the coupling constants and parameters 
from the Ising fixed point to the uniaxial dipolar fixed point.

The combined effect of the uniaxial anisotropy of the exchange interaction and the
anisotropy of the long-range dipolar interaction leads to the following effective free
energy functional~\cite{LK69,Aha73,BZ76,NT76}
\begin{eqnarray}
  {\cal H} = \half \int_{\bf k}
  \left[ r_0 + k^2 + g^2_0 {q^2 \over p^2} \right] 
  \phi_0^{\alpha} ({\bf k}) \phi_0^{\alpha} (- {\bf k})  
  +{u_0 \over 4!} \int_{{\bf x}}  [\phi_0^{\alpha} ({\bf x}) \phi_0^{\alpha} ({\bf x})]^2 \, .
\label{20}
\end{eqnarray}
This problem is characterized by two length scales. Firstly, there is the correlation
length $\xi$, which diverges $\xi \propto r^{-\nu}$ as the critical temperature $T_c$ is 
approached. Here $r$ is the renormalized value of the bare reduced temperature variable
$r_0 = (T-T_c^0)/T_c^0$. Secondly, there is  a length scale introduced by the presence of
the dipolar interaction, given by the so called dipolar wave vector $q_D \propto g_0^2$.

The renormalized parameters, coupling constants and fields are defined by 
$u = \mu^{d-4} Z_u S_d u_0$,
$r = Z_\phi^{-1} Z_r r_0 \mu^{-2}$,
$g = Z_\phi^{-1} Z_g g_0 \mu^{-1}$, and
$\phi_0^{\alpha} ({\bf k}) = Z_{\Phi}^{1/2} \phi^\alpha ({\bf k})$,
where the factor $S_d = {2 / (4 \pi)^{d/2} \Gamma (d/2)}$ is introduced for convenience. 
One finds to one loop order
\begin{equation}
  Z_u =
  1 - u {n+8 \over 6 \varepsilon} 
  \left( 1 + {g} \right)^{-\varepsilon} \, ,
  \quad 
  Z_r = 1 - u {n+2 \over 6 \varepsilon} 
  \left( 1 + { g } \right)^{-\varepsilon} \, ,
\label{21}
\end{equation}
and $Z_{\phi} = Z_g = 1$. The susceptibility can be analyzed in 
terms of the effective exponent 
$\gamma_{\rm eff}={d \ln \chi^{-1}({r},{g},u)/d \ln r}$.
The one loop result is
\begin{equation}
  \gamma_{\rm eff} = -2 /  \zeta_r(g (\ell), u(\ell))
  + {d \ln \chi^{-1} (1,{g (\ell)},u(\ell)) \over d \ln r}
  \, ,
\label{22}
\end{equation}
with the matching condition $r(\ell) / \mu^2 \ell^2  = 1$.
The $\zeta$-function for the mass, is
$\zeta_r(\ell) = -2 + (n+2) u(\ell) ( 1 + g (\ell) )^{-1} / 6 $,
and for $d=3$ the second term in Eq.(\ref{22}) reads
\begin{eqnarray}
&&{d \ln \chi^{-1} (1,{g (\ell)},u(\ell)) \over d \ln r}
  =u(\ell) {n + 2 \over 12} 
  \left( 1 + g^{-2} (\ell) \right)^{-1/2}  \nonumber \\
  &&\quad \quad \quad \quad  \quad \quad \quad  
    \times \biggl[1 + \left( 1 + g^{-1} (\ell)  \right)^{-1}
           - {2 g (\ell) }
                      \ln \left(1 + g^{-1} (\ell)  \right)   
   \biggr] \, .
\label{23} 
\end{eqnarray}
The solution of the flow equation,
$\ell d u(\ell) / d\ell = \beta (\ell)$ with $\beta = \mu \partial_\mu u \big\vert_0$,
for $\varepsilon = 1$ is
\begin{equation}
  u(\ell) = u_{_H} g(\ell) 
  \left[ 
  {g u_{_H} \over u} + 
  \ln \left( {1+g(\ell) \over 1+g} \right) 
  \right]^{-1} \, ,
\label{24}
\end{equation}
where $u_H = 6 \varepsilon / (n+8)$ is the Heisenberg fixed point value.
In the asymptotic limit the susceptibility reduces to    
$\chi^{-1}(r) = r \mid \ln r \mid^{-(n+2)/(n+8)}$
in agreement with Refs.~\cite{Aha73,BZ76}
In Fig.1 the experimental data from Ref.~\cite{FKS82} for
$\gamma_{\rm eff}$  are compared with the theory, where the non universal parameters in
Eqs.(\ref{22})-(\ref{24}) have been chosen as $g=1$ and $u=u_{_H}={1 /3}$ ($n=1$). 
In the range $r \leq 1$ of the reduced temperature there is
excellent agreement between theory and experiment. Especially,
it is found that the observed crossover corresponds to the flow
from the Ising fixed point to the uniaxial dipolar fixed point.
For $r \geq 1$ the data tend to the mean field value
$\gamma_{\rm eff}=1$ corresponding to a crossover
to the Gaussian fixed point. This can be described only
qualitatively within a critical theory by the limit $\ell \rightarrow \infty$.

\begin{figure}
\epsfxsize=3truein
\centerline{\epsffile{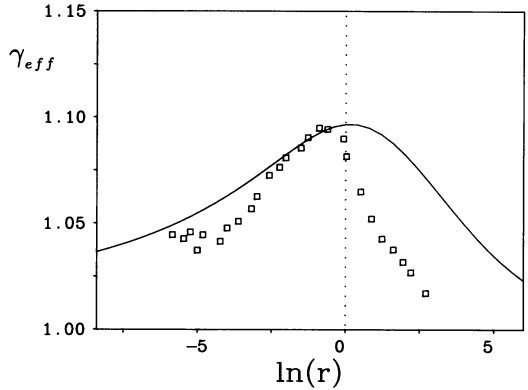}}
\caption{Effective exponent of the susceptibility for 
$g=1.0$ and $u=u_{_I}$ versus $\ln(r)$. The momentum scale is chosen
as $\mu=1$. Experimental data on $LiTbF_4$ from Ref.[20].}
\end{figure}

\subsection{Crossover from isotropic to directed percolation}

Another example, where the presence of a second length scale leads to a crossover in the 
critical behavior, is the physics of biased percolation. The patterns generated by these
percolation models incorporate {\it self-similar} as well as {\it self-affine} clusters.

In ordinary percolation, sites or bonds are filled at random with probability 
$p$~\cite{Ess80}. The percolation process then 
proceeds along paths connecting occupied nearest neighbors. The clusters formed
by nearest-neighbor links are self-similar, i.e., they display isotropic
scaling. In directed percolation \cite{Kin83} the links between nearest
neighbors have a bias in one preferred direction $t$, such that the percolation
process advances along this direction only. The size of the clusters in the preferred
direction is characterized by a length scale different from that in the
perpendicular direction.
If percolation in the positive $t$ direction is merely favored with a certain
probability with respect to the negative $t$ direction, but propagation
``backward'' in time is still admitted, the situation will be more complicated.
If the ``anisotropy'' $g$ is low, one expects almost isotropic scaling behavior
in a large region of the phase diagram. However, when  the critical region 
near the percolation threshold $p_c$ is approached, self-affine scaling will
become apparent.

In order to give a quantitative description of the crossover from
isotropic to directed percolation we investigate the 
pair correlation function $G({\bf r}_2,{\bf r}_1)$, which 
measures the probability that sites ${\bf r}_2 = ({\bf x}_2,t_2)$ and 
${\bf r}_1 = ({\bf x}_1,t_1)$ belong to the same cluster. 
Following the work of Cardy and Sugar~\cite{Car80}, and Benzoni and Cardy~\cite{Ben84},
the pair correlation function (the superscript ``0'' denotes unrenormalized quantities) 
is found to be a sum 
$G^0({\bf r}_2;{\bf r}_1) = \sum_{m,n=1}^\infty [(-i)^{m+n-2} / m! n!]
G^0_{mn}({\bf r}_2;{\bf r}_1)$, where the correlation functions $G^0_{mn}$ are
obtained from the effective action
\begin{eqnarray}
     {\cal J} = \! \int \! \! &&d^Dx \! \! \int \! \! dt 
       \biggl\{ {\tilde \phi}_0({\bf x},t) \!
         \left[ r_0 - {\bf \nabla}^2 - {1 \over c_0^2} \partial_t^2
           + {2 g_0 \over c_0} \partial_t \right] \! \phi_0({\bf x},t) 
       \nonumber \\
       &&+ 
       {u_0 \over 2} \left[ {\tilde \phi}_0({\bf x},t)^2 \phi_0({\bf x},t) - 
       {\tilde \phi}_0({\bf x},t) \phi_0({\bf x},t)^2 \right] \biggr\} \, , 
\label{25}
\end{eqnarray}
with $r_0 \propto p - p_c$.
The effective action is renormalized upon introducing renormalized fields 
$\phi = Z_\phi^{1/2} \phi_0$ and ${\tilde \phi} = Z_\phi^{1/2} {\tilde \phi}_0$,
dimensionless renormalized parameters $r = Z_\phi^{-1} Z_r (r_0 - r_{0c}) \mu^{-2}$, 
$c^2 = Z_\phi Z_c^{-1} c_0^2$, $g = Z_\phi^{-1/2} Z_c^{-1/2} Z_g g_0 \mu^{-1}$, and a
renormalized coupling constant 
$u = Z_\phi^{-3/2} Z_u u_0 B_d^{1/2} \mu^{(d-6)/2}$.
$B_d = \Gamma(4-d/2) / (4 \pi)^{d/2}$ is a geometric factor,
and $r_{0c}$ denotes the fluctuation-induced shift of the percolation
threshold. The five independent
multiplicative renormalization constants are then uniquely determined by
extracting the ultraviolet singularities of the two- and three-point vertex
functions at finite mass $r_0 = \mu^2$. In the course of renormalizing the model it is
convenient to introduce an effective anharmonic coupling\footnote{Here 
$I^d_{mn}(g) = \int_0^1 {x^{m/2 - 1} (1 + x g^2)^{(d-m-n)/2}} dx $ (with
odd integers $m$ and $n$) denotes a certain type of integrals, with
$I^d_{mn}(0) = 2/m$, and $\lim_{g \rightarrow \infty} [g^m I^d_{mn}(g)] =
\Gamma(m/2) \Gamma((n-d)/2) / \Gamma((m+n-d)/2)$.} $v = u^2 c I^d_{17}(g)$.

From an analysis of the renormalization group equation (where for 
notational convenience we have taken $\mu=1$ in the following)
\begin{equation} 
     \Gamma_{11}(\{ x \}, {\bf q},\omega) =                 
     \ell^2 e^{ \int_1^\ell \zeta_\phi(\ell') d\ell' / \ell' }
       {\hat \Gamma}_{11} \left( r(\ell), v(\ell), {{\bf q} \over \ell},
                           {g(\ell) \omega \over c(\ell) \ell},
                      {\omega^2 \over c(\ell)^2 \mu^2 \ell^2} \right) \, .
\label{26}
\end{equation} 
one finds {\it four} independent critical exponents. At a fixed point $v^*$ using the 
matching condition ${\ell = q }$, one arrives at the following general 
{\it self-affine} scaling form
\begin{equation}
     \Gamma_{11}(r,c,g,u,{\bf q},\omega) \propto q^{2 - \eta_\perp}
      {\hat \Gamma}_{11} \left( {r \over q^{1 / \nu_\perp}}, v^*, 1,
                          {g \omega / c \over q^z},
      {\omega^2/ c^2 \over q^{2 z (1 - \Delta)}} \right) \, , 
\label{27}
\end{equation}
with the critical exponents
$\eta_\perp = - \zeta_\phi^*$, $\nu_\perp = - 1 / \zeta_r^*$, $z = 1 +
\zeta_c^* - \zeta_g^*$, and $z \Delta = - \zeta_g^*$. Here, $\eta_\perp$ and
$\nu_\perp$ correspond to the two independent indices familiar from the theory
of static critical phenomena. The exponent $z$ was introduced in analogy to a
dynamical critical exponent, and in our case is related to the anisotropic
scaling behavior. Finally, $\Delta$ is a positive crossover exponent describing
the transition from isotropic to directed percolation. It stems from the fact
that there appear {\it two} different scaling variables for the ``frequency''
$\omega$ in Eq.~(\ref{27}). In the asymptotic limit of directed percolation, $g
\rightarrow \infty$, the second scaling variable disappears, and the scaling
behavior is described by the {\it three} exponents $\eta_\perp$, $\nu_\perp$,
and $z$.

Similarly, with the choices $\ell = \left( g \omega / c \right)^{1 / (1 +
\zeta_c^* - \zeta_g^*)}$ and $\ell = r^{-1 / \zeta_r^*}$ the longitudinal exponents
$2 - \eta_\parallel = (2 - \eta_\perp)/z$ and $\nu_\parallel = z \nu_\perp$
and the (``susceptibility'') exponent $\gamma = \nu_\perp (2 - \eta_\perp) = 
\nu_\parallel (2 - \eta_\parallel)$, respectively. 
In the special case of isotropic percolation ($g_0 = 0$), the
scaling relations simplify considerably, describing {\it self-similar} scaling
with only {\it two} independent critical exponents $\eta = - \zeta_\phi^*$ and
$\nu = -1 / \zeta_r^*$.

We have seen that both the self-similar and the self-affine
scaling behavior are within the scope of the present theory, at least for
dimensions $d < 5$; for $5 < d \leq 6$ the model is not renormalizable in the
directed limit, and simply characterized by the exponents corresponding to the
Gaussian fixed point $v^*_{\rm GD}$, with logarithmic corrections for $d = 5$.
This again emphasizes the fact that no expansion with respect to a fixed upper
critical dimension can be applied consistently. In order to study the entire 
crossover region between these asymptotic regimes, the coupled set of flow 
equations have to be solved 
numerically. Fig.~2 shows the flow diagram (at $d = 3$ dimensions) for the 
effective couplings $v(\ell)$ and ${\bar g}(\ell)$, whose topology is 
determined by the four fixed points. The separatrix connecting
$v^*_{\rm I}$ with the infrared-stable fixed point $v^*_{\rm D}$ describes the
universal crossover features from isotropic to directed percolation.

\begin{figure}
\epsfxsize=3truein
\centerline{\epsffile{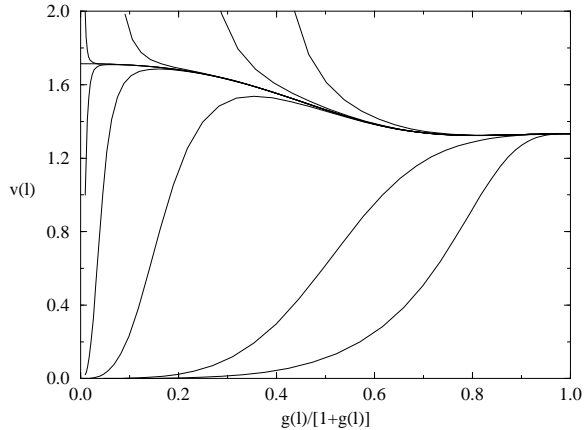}}
\caption{Renormalization-group trajectories in the $(v,{\tilde g})$ parameter 
space for the crossover from isotropic to directed percolation (to one-loop
order) in $d = 3$ dimensions.}
\end{figure}

The interchange from self-similar to self-affine scaling is most conveniently
described by introducing effective exponents for the pair-correlation function.
Using the zero-loop result for the two-point vertex function
\begin{eqnarray}
     \Gamma_{11} ( \{ x \} ,{\bf q},\omega) = \ell^2
       e^{\int_1^\ell \zeta_\phi(\ell') d \ell'/ \ell'} 
     \left[ r(\ell) + \left( {q \over \ell} \right)^2 +
                {\omega^2 \over \ell^2 c(\ell)^2} + 
               2 i {\omega g (\ell) \over \ell c(\ell)} \right] \, , 
\label{28}
\end{eqnarray}
we specialize to $r = 0$ and ${\bf q} = {\bf 0}$ and define
\begin{equation} 
     2 - \eta_{\parallel \, \rm eff}(\omega) = 
      {d \ln {\sqrt{\mid \Gamma_{11}(0,{\bf 0},\omega) \mid^2 }}
       \over d \ln \omega}       \quad . 
\label{29}
\end{equation}
Using $\left \vert \omega^2 / \ell^2 c(\ell)^2 + 2 i \omega g(\ell) / 
\ell c(\ell) \right \vert^2 = 1$ yields $2 - \eta_{\parallel \, \rm eff}(\ell)
= [ 2 + \zeta_{\phi}(\ell) ] d \ln \ell / d \ln \omega$. Similarly, in the case
$r = \omega = 0$ one finds
\begin{equation} 
     2 - \eta_{\perp \, {\rm eff}}(q) = { d \ln \Gamma_{11}(0,{\bf q},0)
      \over d \ln q} \, , 
\label{30}
\end{equation}
which reduces to $2 - \eta_{\perp \, {\rm eff}} = 2 + \zeta_\phi(\ell)$, if $(q
/ \ell)^2 = 1$ is inserted. Finally, considering ${\bf q} = {\bf 0}$ and
$\omega = 0$ one introduces
\begin{equation} 
    \gamma_{\rm eff}(r) = {d \ln \Gamma_{11}(r,0,0) \over d \ln r} \quad ,
\label{31}
\end{equation} 
and choosing the matching condition $r(\ell) = 1$ we find $\gamma_{\rm
eff}(\ell) = - [2 + \zeta_\phi(\ell)] / \zeta_r(\ell)$.

The flow of the effective exponents $\eta_{\parallel \, \rm eff}(\ell)$,
$\eta_{\perp \, \rm eff}(\ell)$, and $\gamma_{\rm eff}(\ell)$ in $d = 3$
dimensions is depicted in Fig.~3, with the initial value
for the coupling $v(1) = v^*_{\rm I}$ of the isotropic scaling fixed point. The
dependence on the anisotropy scale $g$ was eliminated by plotting versus the
scaling variable $\ln g(\ell)$; the graphs corresponding to different initial
values $g(1)$ then all collapse onto one master curve. The most important
conclusion to be drawn from Fig.~3 is that the anisotropy scale, at which the
crossover occurs, considerably differs for the effective exponents defined
above. $\eta_{\parallel \, \rm eff}$ starts to cross over from the isotropic
to the directed fixed point value already at $\ln g(\ell_{\rm cross}) \approx
-0.8$, whereas $\gamma_{\rm eff}$ shows this crossover at $\ln g(\ell_{\rm
cross}) \approx -0.2$, and $\eta_{\perp \, \rm eff}$ only at $\ln g(\ell_{\rm
cross}) \approx +0.8$. Note that the sizeable change of $\eta_{\parallel \, \rm
eff}$ is already apparent at mean-field level, where it acquires the values $0$
and $1$ in the isotropic and directed limit, respectively. However, a crossover
of the exponents $\eta_{\perp \, \rm eff}$ and $\gamma_{\rm eff}$ requires the
$\zeta$ functions at least on the one-loop level.

\begin{figure}
\epsfxsize=3truein
\centerline{\epsffile{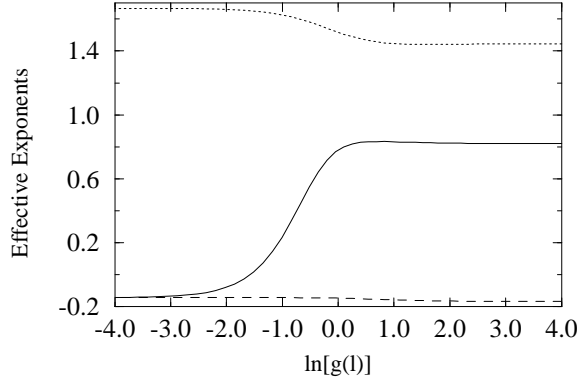}}
\caption{Effective exponents $\eta_{\parallel \, {\rm eff}}(\ell)$ (solid),
$\eta_{\perp \, {\rm eff}}(\ell)$ (dashed), and $\gamma_{\rm eff}(\ell)$
(dotted) for the connectivity as functions of the running anisotropy parameter
$g(\ell)$ for a fixed initial value of the coupling constant $v(1) = v^*_{\rm I}$.}
\end{figure}

\subsection{Multicritical behavior in lattice boson models}

The critical behavior at the zero temperature Mott insulator to superfluid
(S-I) transition of lattice boson models can be analyzed in terms of the effective 
action~\cite{FWG89}
\begin{equation}
  {\cal S} = \half \int_{{\bf k}, \omega}
  \left[ r_0 + k^2 + \omega^2 + 2 ig_0 \omega \right] \mid  \phi_0 ({\bf k},\omega )\mid^2
 + {u_0 \over 4} \int_{{\bf x}, t} \mid  \phi_0 ({\bf x},t)\mid^4 \, ,
\label{32}
\end{equation}
where the field $\phi$ serves as the order parameter for superfluidity. The parameters
$r_0$ and $u_0$ can be calculated from the underlying microscopic lattice models. 
One finds that $g_0(\mu, J) = \partial r_0(\mu, J) / \partial \mu$, where $\mu$ and $J$ 
are the chemical potential and the hopping matrix element of the lattice model, 
respectively. Since in mean field theory the phase boundary is given by $r_0(\mu, J) = 0$,
the point $g_0=0$ constitutes a multicritical point for S-I transition.

Before starting the renormalization group calculation of the crossover scaling
functions, we give a brief review of the phenomenological scaling
theory~\cite{FWG89}.The quantum phase transition at $T=0$ is characterized by both a
diverging length scale $\xi \propto r^{-\nu}$ and a diverging time scale 
$\Omega^{-1} \propto r^{-\nu z}$. Near the multicritical point $(r,g)=(0,0)$ the
singular part of the free energy shows crossover scaling behavior
\begin{equation}
  f_s \propto r^{(d+z)\nu} {\tilde f}_s (g/r^\phi) \, ,
\label{33}
\end{equation}
with $\phi$ being the crossover scaling exponent. As discussed in section 2 the
crossover scaling function ${\tilde f}_s (x)$ has to become singular at some critical 
value $x_c$ in order to describe the crossover from primary to secondary critical 
exponents.

Here we use the field theoretic method described in section 2 to determine the effective 
exponents. We introduce renormalization factors by $\phi = Z_\phi^{1/2} \phi_0$, and 
dimensionless renormalized parameters $r = Z_\phi^{-1} Z_r (r_0 - r_{0c}) \mu^{-2}$, 
$g = Z_\phi^{-1/2} Z_g g_0 \mu^{-1}$, and a renormalized coupling constant 
$u = Z_\phi^{-2} Z_u u_0 B_d \mu^{d-3}$. The crossover induced by the term
linear in the frequency $\omega$ leads to a reduction of the upper critical dimension from
$d_c = 3$ to $d_c = 2$. From a one-loop calculation we find~\cite{Fre95} for the
renormalization factors $Z_\phi = 1$, $Z_g = 1$, and
\begin{eqnarray}
  Z_r &&= 1 - {u \over \varepsilon} \int_o^1 dx x^{m/2-1} (x+g^2)^{(d-m-n)/2} \, , \\
  Z_u &&= 1 - {1 \over \varepsilon} {\bar A} u - {1 \over \varepsilon} {\bar B} u \, ,
\end{eqnarray}
where ${\bar A} = \quart I_{12}^d (g)$ and ${\bar B} = (1+g^2)^{\varepsilon/2}$ with
$\varepsilon = 3-d$ being the distance to the upper critical dimension of the primary fixed
point. One should note that $Z_g = 1$ is valid to all orders in perturbation theory since
there is no singular contribution proportional to $\omega$. The one-loop theory may be
further improved by taking into account results which become exact in the limit $g
\rightarrow \infty$. In this limit the special (``causal'') structure of the propagator
implies that the corrections to the bare two-point vertex function vanish to all orders
in perturbation theory~\cite{Uzu81,FWG89}, i.e., $Z_r (g \rightarrow \infty) = 1$
exactly. Furthermore, there are only ``ladder''-diagrams~\cite{Uzu81} contributing to the
renormalization of $u$, which form a geometric series with the result
\begin{equation}
  u \mu^\varepsilon = {u_0 \over 1 + A u_0} \, ,
\end{equation}
with $A = {1 \over \varepsilon} B_d \mu^{-\varepsilon} {\bar A}$. Now, in order to improve the
one-loop result we suggest to resum the ladder diagrams corresponding to the one-loop
term $A u_0$, while still using the one-loop result for all the other terms. This
procedure amounts to
\begin{equation}
  u \mu^\varepsilon = {u_0 \over 1 + A u_0} - B u_0 \, ,
\end{equation}
where $B = {1 \over \varepsilon} B_d \mu^{-\varepsilon} {\bar B}$. Equivalently to one loop
order we could also use   $u \mu^\varepsilon = u_0 / ( 1 + (A+B) u_0)$.  

Upon defining Wilson's flow functions as in section 3.2 one can study the behavior of the
vertex function $\Gamma_{11}$ under renormalization group transformations
\begin{equation}
  \Gamma_{11}(\mu, r, g, u; q, \omega) =
  \mu^2 \ell^2 {\hat \Gamma}_{11} \left( r(\ell), g(\ell), v(\ell); {q \over \mu \ell},
                                         {\omega \over \mu \ell} \right) \, ,
\end{equation}
where it is convenient to introduce the effective coupling constant $v = u I_{12}^d (g)$.
At the secondary fixed point $v^*$ one finds using the matching condition $r(\ell) = 1$
\begin{equation}
  \Gamma_{11}(\mu, r, g, u; q, \omega) \propto  \ell^{2 \nu}
  {\hat \Gamma}_{11} (1, g \xi, v^*; q \xi, \omega \xi ) \, ,
\end{equation}
where $\xi \propto r^{- \nu}$. Since $Z_g = 1$ to all orders in perturbation theory, one
finds that the crossover exponent $\phi$ is exactly given by the correlation length
exponent. One can define an effective exponent $\nu(\ell) = -1 / \zeta_r (\ell)$, the
flow of which is obtained from a solution of the flow equations~\cite{Fre95}. An
effective dynamic exponent may be obtained by analyzing the pole $\Gamma$ of the
dynamic susceptibility (we have used the 0--loop result for the scaling function)
\begin{equation}
 \Gamma \propto \mu \ell \left[ g(\ell) - \sqrt{r(\ell) + g^2(\ell) + 
                                q^2 / (\mu \ell)^2 }  \right] \, ,
\end{equation}
the scaling behavior of which may be analyzed in various ways~\cite{Fre95}. 
At the critical point $r=0$ with $q/ \mu \ell = 1$ we get
\begin{equation}
  \Gamma \propto q \left[ g/q- \sqrt{1 + g^2 / q^2 } \right] 
\end{equation}
describing the trivial crossover from $z=1$ to $z=2$ as $q \rightarrow 0$. Note, that
there is, however, nontrivial crossover behavior for the exponent of the correlation
length from the Ising fixed point value $\nu^*_I$ to the Gaussian fixed point value
$1/2$.

\section{Summary and outlook}
In this paper a field theoretic renormalization group technique using 
normalization conditions was presented, which allows for the description of crossover
phenomena, where the primary and secondary fixed point have different upper critical
dimension. This method can be thought of as the field theoretic version of Nelson's
trajectory integral method. It has the advantage of being capable of describing quite
complicated crossover scenarios. Especially, using the field theoretic renormalization
method one can derive an explicit expression for the crossover scaling function, where
the regular and singular parts are already separated.

We have employed this method for three particular interesting crossover phenomena from
various fields of critical phenomena, namely magnetic phase transitions with dipolar
anisotropy, biased percolation problems, and the multicritical point of the Mott
insulator to superfluid transition in Bose lattice models.

Another variant of Amit and Goldschmidt's generalized minimal subtraction procedure has
been used by Stephens and O'Connor~\cite{Crs,Crs94} 
to investigate crossover phenomena ranging from finite size scaling problems to
quantum-classical crossover.

We close with emphasizing that the method could be applied to various other quite
interesting crossover phenomena. Since it is quite likely that many experiments
are not done in the asymptotic but in the crossover regime, it seems highly desirable
to perform such calculations in order to get a quantitative understanding of the
critical behavior.

\ack{It is a pleasure to thank U.C. T\"auber and F. Schwabl for collaboration on some of
the subjects discussed in this paper. I also would like to acknowledge financial support
from the Deutsche Forschungsgemeinschaft under contract no. Fr 850/2.}

\end{document}